# Raman fingerprints of ultrasmall nanodiamonds produced from adamantane

Oleg Kudryavtsev[1], Rustem Bagramov[2], Arkady Satanin[3,4], Oleg Lebedev[5], Dmitrii Pasternak[1], Vladimir Filonenko[2], Igor Vlasov[1*]

[1] Prokhorov General Physics Institute of the Russian Academy of Sciences, 119991 Moscow, Russia; vlasov@nsc.gpi.ru; leolegk@mail.ru
[2] Vereshchagin Institute of High Pressure Physics, Russian Academy of Sciences, 108840 Moscow, Troitsk, Russia; bagramov@mail.ru; vpfil@mail.ru
[3] Dukhov All-Russia Research Institute of Automatics, 101000 Moscow, Russia; asatanin@gmail.com
[4] National Research University Higher School of Economics, 101000 Moscow, Russia; asatanin@gmail.com;
[5] Laboratoire CRISMAT, UMR 6508 CNRS-ENSICAEN, 14050, Caen, France; oleg.lebedev@ensicaen.fr
* Correspondence: vlasov@nsc.gpi.ru

**Abstract:** The synthesis of ultrasmall (2-5 nm) nanodiamonds purely from adamantane at pressure of 12 GPa is reported. Their structural features have been studied by Raman spectroscopy. The unusual vibration band containing a number of pronounced maxima at about 1147, 1245, 1344, and 1456 cm$^{-1}$ was detected in Raman spectra. The band is confidently identified with the bending vibrational modes of CHx groups terminating the nanodiamonds surface. Excessively intense mode at 1344 cm$^{-1}$ is explained by its coupling with the 1328 cm$^{-1}$ diamond phonons. The Raman band found is proposed to be used for express recognition of ultrasmall nanodiamonds produced from adamantane and other hydrocarbons with a high hydrogen content. Moreover, polarized CH bonds on a diamond surface are sensitive to environmental conditions. This opens up opportunities for using the diamond produced from adamantane as ultrasmall nanosensors in biology, chemistry, and medicine

## 1. Introduction

Until recently, researchers had at their disposal two main types of ultrasmall (smaller than 10 nm) diamond particles of bottom-up synthesis produced by detonation and CVD methods. Both types are easily recognizable using the express and non-destructive technique of Raman spectroscopy.

The first diamond nanoparticles smaller than 10 nm in size (hereinafter referred to as ultrasmall nanodiamond or UND) were synthesized by detonation technique more than half a century ago [1, 2]. Since then the detonation nanodiamonds have found application in various scientific and technological fields: tribology, drug delivery, bioimaging and tissue engineering [3–5]. They are easily recognizable using Raman spectroscopy. Due to the smallness of their characteristic size (usually about 3-5 nm) and even smaller sizes of coherent scattering regions within one crystallite, the phonon confinement effect [6] is clearly manifested there, downshifting the Raman diamond line relative to the 1332.5 cm$^{-1}$ position characteristic of bulk diamond. In the spectrum of detonation UND, well purified from amorphous carbon, particular line is observed in the region of 1630–1650 cm$^{-1}$, the origin of which is still a subject of debate [7–9]. Since this line is not observed in UND synthesized by other techniques, it can be considered as Raman fingerprint of detonation nanodiamonds.

In the late 1990s, another ultrasmall nanodiamonds were synthesized by CVD technique. They formed polycrystalline film and had characteristic sizes of 2–5 nm [10–11]. The name "ultra-nanocrystalline diamonds" (UNCD) stuck behind those nanostructures. In the Raman spectrum of UNCD, in addition to the main diamond line, two unusual lines were present at 1140 cm$^{-1}$ and 1490 cm$^{-1}$.

Comparing the Raman spectra of UNCD and trans-polyacetylene recorded at different excitation wavelengths they concluded that these lines are associated with bending vibrations of trans-polyacetylene (tPA) chains, apparently formed at diamond intergrain boundaries [12]. The lines at 1140 cm$^{-1}$ and 1490 cm$^{-1}$ are considered to be Raman fingerprints of ultra-small polycrystalline diamonds grown by CVD method.

Bottom-up synthesis of a new generation of nano- and micro-diamonds from various hydrocarbons at high pressure and high temperature (HPHT) began to be mastered recently [13, 14, 15]. The synthesis of diamond was carried out at a pressure of 8-9 GPa in the temperature range of 1000-1700$^0$C, and the size of the resulting diamond decreased with decreasing synthesis temperature. Note, that the stable synthesis of diamond purely from hydrocarbons (like adamantane) at such pressures occurred only at temperatures >1300$^0$C and the characteristic size of diamond nanoparticles exceeded 10 nm. The synthesis temperatures below 1300$^0$C initiate a competing transformation of hydrocarbons into graphite [15,16]. The addition of various halogens to the hydrocarbon growth medium made it possible to lower the threshold temperature for the stable growth of diamond at a pressure of 8 GPa and produced an ultrasmall nanodiamond in the size range of 1–10 nm [17, 18]. In the Raman spectra of UNDs obtained from halogenated adamantane, new peaks were observed in the frequency band 1000-1500 cm$^{-1}$, which were not previously detected in other UNDs produced by detonation and CVD methods. However, the identification of those peaks was ambiguous, they were first attributed to halogens adsorbed on a diamond surface [17], then to C=C bonds on a diamond surface and to diamond phonon density of states [18]. Only one of those peaks was ascribed to CH bending mode [18]. In our opinion, the main structural features revealed by Raman spectroscopy in nanodiamonds produced from adamantane and other hydrocarbons should be related to the extremely high concentration of hydrogen terminating the diamond surface [19, 20]. Reliable confirmation of our hypothesis required the synthesis of UND purely from adamantane. We have succeeded in producing such a UND using a pressure of 12 GPa and a temperature of about 1300$^0$C. The choice of synthesis conditions was based on the thermodynamic model, according to which hydrocarbons at pressure > 10 GPa and temperature > 1000 K dissociate into diamond and molecular hydrogen [21]. Despite the fact that the possibility of synthesizing nanodiamonds from admantane at 12 GPa was reported quite a long time ago by Wentorf [22], as far as we know, experiments with adamantane at this pressure have not been repeated since then, and the nanodiamond synthesized in this way has not yet been studied using Raman spectroscopy. In this paper, we present the results of a structural analysis of the UND using high-resolution transmission electron microscopy and Raman spectroscopy.

## 2. Results and discussion

In the present work, we reproduced the classical experiment [22] on the thermobaric treatment of hydrocarbons at a pressure of 12 GPa. Adamantane C10H16 (99%, Aldrich) was used as a precursor. High pressures were provided by toroid type equipment. The details of the diamond synthesis are given in Methods.

The characteristic sizes of synthesized diamond crystallites of 2-5 nm were determined by high-resolution transmission electron microscopy (HRTEM) (Fig.1) and Raman spectroscopy (see below). Based on numerous HRTEM images the diamond grains were concluded to be well crystallized, do not contain structural defects in the form of twins and have a predominantly rounded shape (Fig. 1b,c).

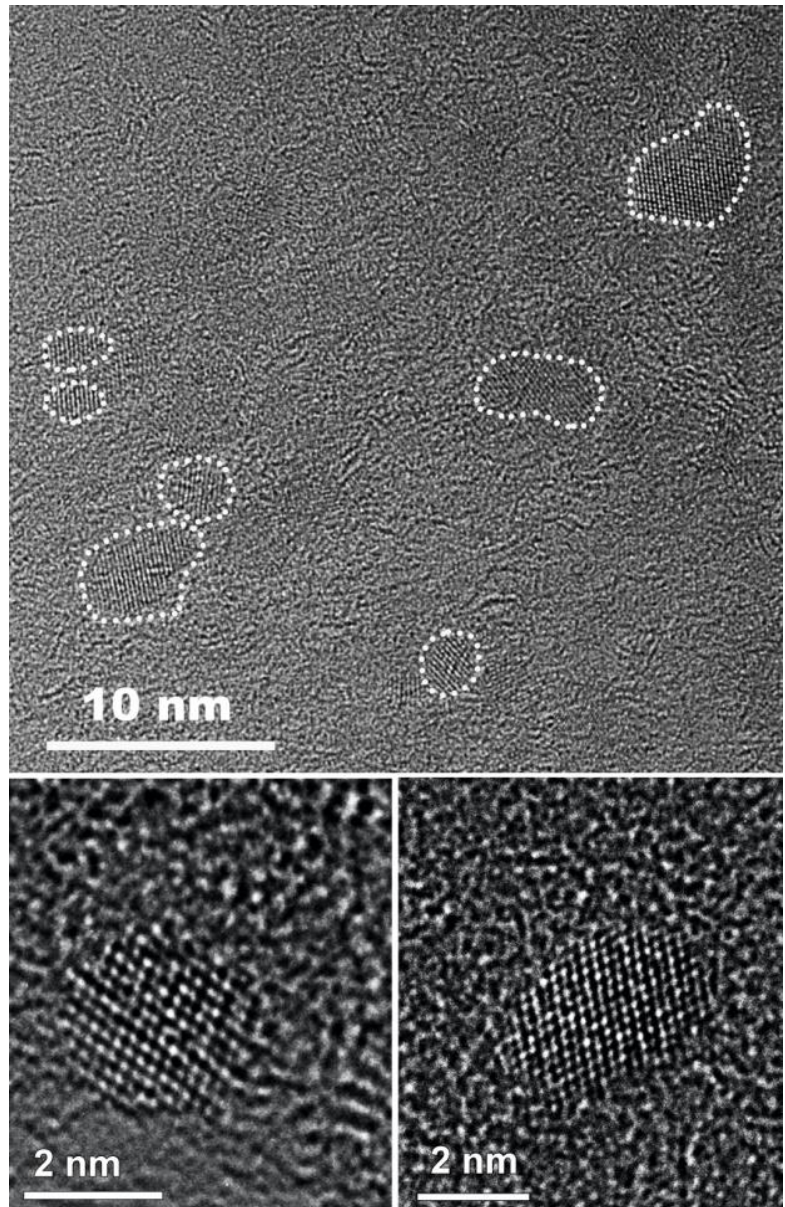

**Figure 1.** (a) HRTEM image of typical diamond crystallites, depicted with dotted lines. Particle sizes range from 2 to 5 nm. (b, c) Enlarged HRTEM images of single diamond particles of spherical and oval shapes.

The characteristic Raman spectrum of the UNDs recorded at 473-nm laser excitation is shown in Figure 2. Two distinct vibrational bands are observed in the frequency ranges of 1000-1500 cm$^{-1}$ and 2800-3000 cm$^{-1}$. The higher frequency range is typical for CHx (x=1,2,3) stretching vibrations observed in the Raman and IR absorption spectra of various organic molecules [23], diamondoids [24,25] and nanodiamonds of various sizes subjected to post-synthesis hydrogenation [26].

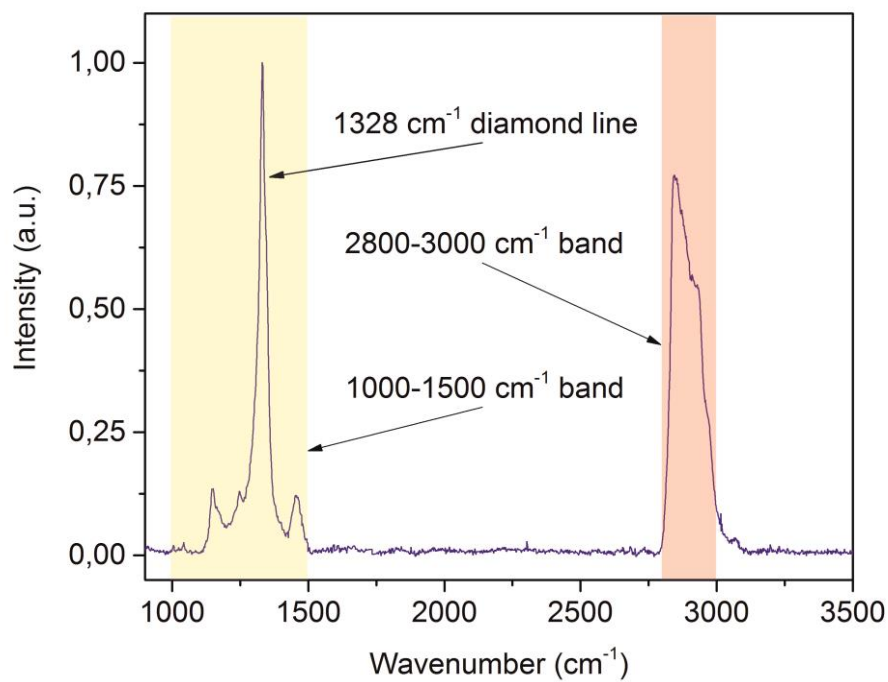

**Figure 2.** Raman spectrum of the sample recorded at 473-nm laser excitation. Two distinct vibrational bands are observed in the frequency ranges 2800-3000 cm$^{-1}$ and 1000-1500 cm$^{-1}$. The spectrum is shown after subtraction of a small luminescent background and normalization to the diamond line intensity.

Now consider Raman spectra recorded at different laser excitation wavelengths in the range 1000-1500 cm$^{-1}$ (Fig. 3). Raman-active phonon mode of diamond is observed at 1328 cm$^{-1}$. The frequency shift relative to the 1332.5 cm$^{-1}$ (position of bulk diamond line) is due to the phonon confinement effect, and the shift value is determined by the diamond size [6,27]. A close frequency shift down to 1328.5 cm$^{-1}$ is also observed for Raman line of detonation nanodiamonds, whose characteristic sizes, according to HRTEM, are 3–6 nm [28]. Thus, the Raman (Fig. 3) along with TEM analysis confirms that the typical sizes of the UND studied in this work are below 6 nm. A wide band, unusual for nanodiamonds, is superimposed on the diamond line. The intensity of this band at maxima is about an order of magnitude weaker than that of the diamond line. The most pronounced and intense peaks of the band are positioned at about 1147, 1245, and 1456 cm$^{-1}$. No frequency dispersion of these peaks occurs with a change in excitation energy (Fig 3a and 3b). This fact rules out the identification of these peaks with the vibrational modes of transpolyacetylene, which are typically detected in CVD UND polycrystallites in the range 1000–1500 cm$^{-1}$ [12].

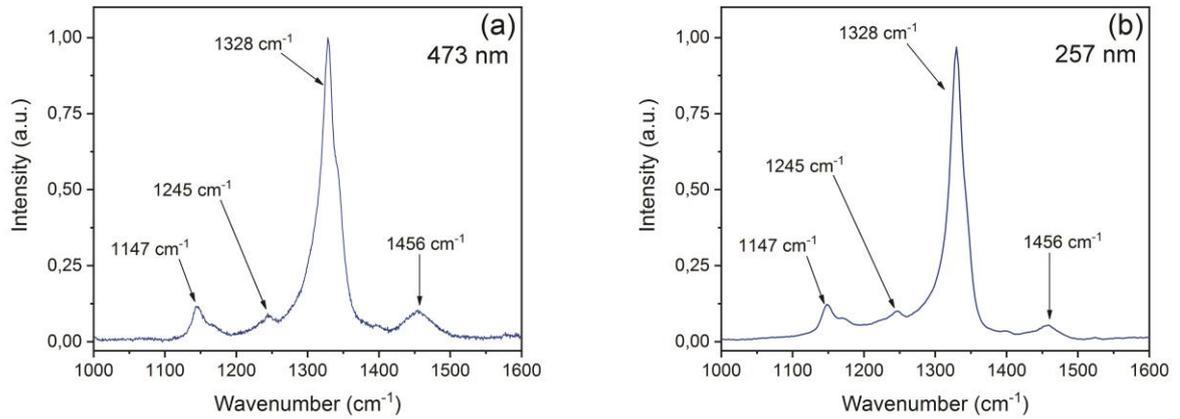

**Figure 3.** Raman spectra (black curves) recorded in the frequency range 1000-1500 cm$^{-1}$ under laser excitation at 473 nm (a) and 257 nm (b). The strong line at 1328 cm$^{-1}$ are related to diamond phonon modes. The most intense peaks at 1147, 1245, 1456 cm$^{-1}$ are ascribed to CHx bending modes.

Further, take a closer look at the spectrum around the diamond line (Fig. 4). Deconvolution of the spectrum in the range 1250-1400 cm$^{-1}$ into Lorentz profiles gives 3 components. The component at 1312 cm$^{-1}$ we associate mainly with the phonon modes of UNDs. Two components at 1312 cm$^{-1}$ and 1328 cm$^{-1}$ form an asymmetric profile with an extended low-frequency shoulder. A similar diamond line is also observed in Raman spectra the detonation UNDs. The appearance of asymmetry in a diamond line shape with tailing towards lower frequencies is explained by a phonon confinement effect [6,27]. The component at 1344 cm$^{-1}$ belongs to a set of unusual peaks seen the range 1000-1500 cm$^{-1}$, and is the most intense among them. It is only 3 times less intense than the diamond line, while the other peaks are 10 or more times weaker than the diamond line. A possible reason for this could be an interaction of 1344 cm$^{-1}$ vibrational mode with the diamond phonons due to their spectral overlapping. A simple generalization of Placzek's theory [29] to the case of two coupled optical oscillators (with slightly different Raman tensors) explains a possible amplification of the 1344 cm$^{-1}$ mode due to the energy "pumping" from the more intense 1328 cm$^{-1}$ mode (see SI for details). An example of Raman spectra simulated for two modes close in frequency is shown in Figure 5. Different colors show partial and total contributions from the uncoupled oscillators and from the coupling ones calculated at coupling coefficient k=90 cm$^{-1}$. Switching-on the interaction increases an oscillation intensity at the frequency of 1344 cm$^{-1}$.

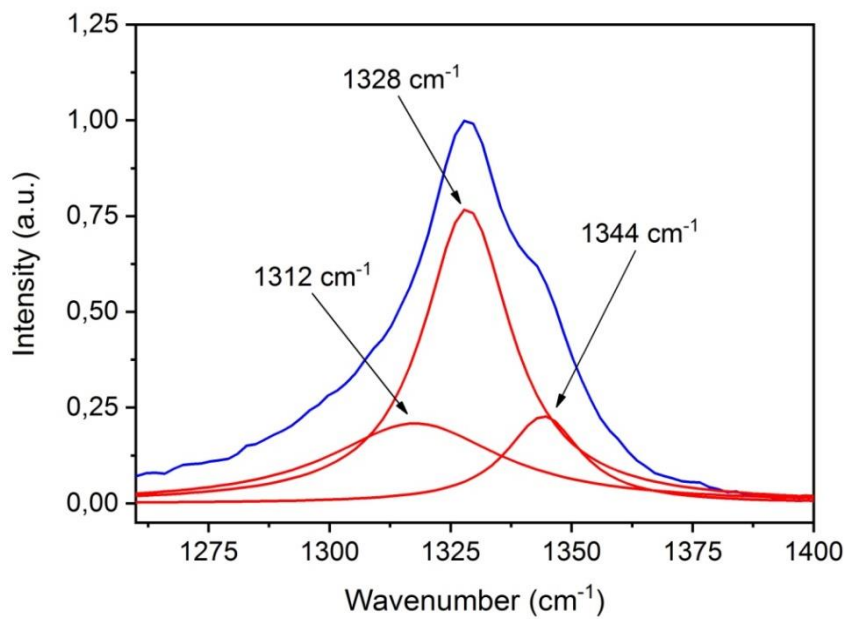

**Figure 4.** Raman spectrum in the range 1250-1400 cm$^{-1}$. Deconvolution of the spectrum into Lorentz profiles are shown in red. The components at 1312 cm$^{-1}$ and 1328 cm$^{-1}$ refer to diamond phonon modes, at 1344 cm$^{-1}$ – to CHx bending modes.

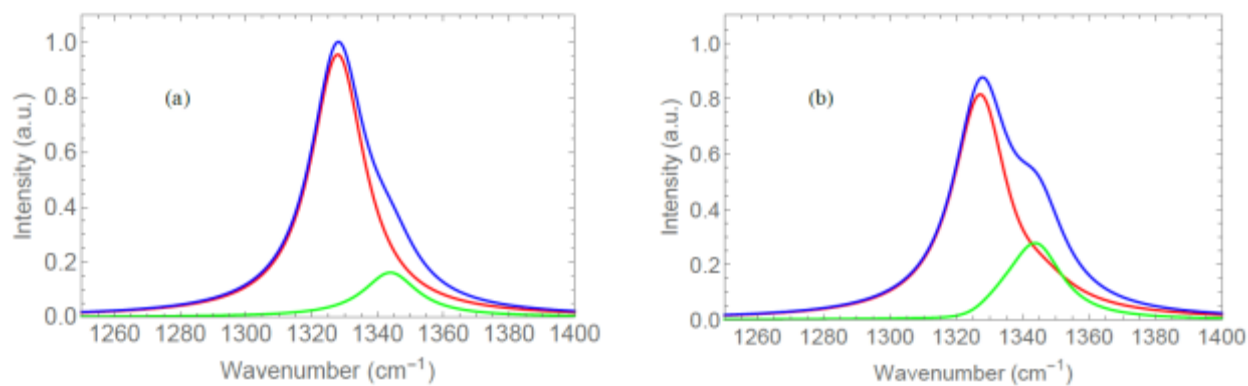

**Figure 5.** Numerical calculation of the Raman spectrum of two modes close in frequency. (a) two uncoupled modes: red - partial contribution of the 1328 cm$^{-1}$ mode; green - partial contribution of the 1344 cm$^{-1}$ mode; blue - total contribution. (b) Two coupled modes; the coupling coefficient is 90 cm$^{-1}$. On (b) one can see the effect of amplification of the 1344 cm$^{-1}$ mode due to the "pumping" of energy from the more intense 1328 cm$^{-1}$ mode.

CHx bening vibrations of hydrocarbon molecules [23] and diamondoids of various sizes [24] are known to be localized in the 1000-1500 cm$^{-1}$ region. Similar to the range 2800-3000 cm$^{-1}$ of CHx stretching vibrations common for various hydrocarbon molecules, diamondoids and nanodiamonds, it is logical to assume a common range of CHx bending vibrations for these materials. Thus, the Raman band with maxima at 1147, 1245, 1344, 1456 cm$^{-1}$ are most likely associated with bending vibrations of CHx groups. For diamond particles of 2–5 nm in size, the number of carbon atoms on the surface exceeds 15%. Assuming that each surface C atom is terminated by at least one hydrogen atom, we find that the H/C ratio in the studied UND is not much inferior to that in large diamondoids, in particular, in cyclohexamantane ($C_{26}H_{30}$). Since the CHx bending modes are detected in the Raman spectrum of $C_{26}H_{30}$ [30], it is reasonable to expect similar Raman modes for the strongly hydrogenated UND. No sp$^2$-bound carbon was found in the nanodiamond sample under study, therefore, the presence of any other vibrational modes in addition to diamond and CHx modes in nanodiamond produced purely from adamantane, is unlikely. This again confirms our assertion that all unusual lines in the range 1000-1500 cm-1 are related to the CHx modes. To our knowledge, no observation of CHx bending modes in the Raman spectra of any other types of nanodiamonds (even after their special treatment with hydrogen) has been reported so far. We explain this fact by the low degree of surface hydrogenation as well as by the lower intensity of the CHx bending modes compared to the stretching ones. An experimental confirmation of the high degree of surface hydrogen termination for our sample is the unusually high intensity of CHx valence modes, comparable to the intensity of the diamond mode in Raman (Fig. 2).

Experiments on UND annealing in air confirm the relationship between the 1000-1500 cm$^{-1}$ band and hydrogen on a diamond surface. The sample was annealed sequentially at 200, 300, and 400 $^{o}$C (see Methods). Annealing temperatures of 200 and 300 $^{o}$C do not notably change its Raman spectrum, while upon annealing at 400°C, the intensities of the bands 1000-1500 cm$^{-1}$ and 2800-3000 cm$^{-1}$ 1 decrease essentially with respect to the diamond line (Figure 6). All the main peaks of the first band become practically indistinguishable against spectral noises. The integral intensity of 2800-3000 cm$^{-1}$ band decreases by 4 times, but this occurs unevenly within the band (Fig. 6b). The intensity in the range 2800-2900 cm$^{-1}$ dropped more than that in the range 2900-3000 cm$^{-1}$. We explain such a difference by the faster dissociation of lower-energy CHx bonds. As the energy (frequency) that characterizes the stretching vibration of atomic bonds is proportional to the bond dissociation energy [31], it is reasonable to observe a predominant intensity reduction for CHx modes of lower frequency.

In addition to changes in the Raman spectra, associated with desorption of hydrogen from the diamond surface, annealing at 400°C reduces the diamond line intensity by about 2 times (Fig. 6a,b), and strong broadband luminescence appears. We attribute the weakening of the diamond line at 1328 cm$^{-1}$ to the partial removing oxidized carbon from the sample. The emerging luminescence is a feature of oxidized nanodiamonds [32,33].

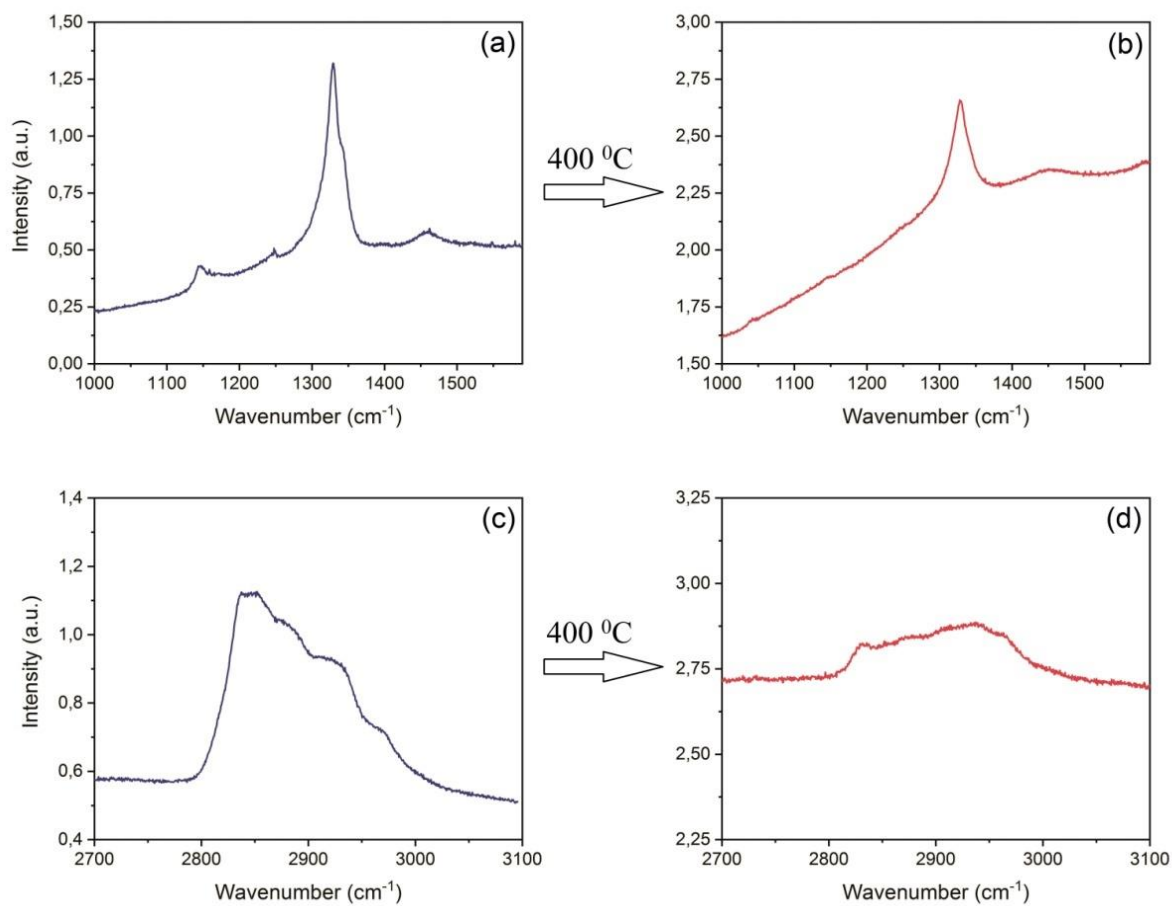

**Figure 6.** Raman spectra recorded in the ranges 1000-1600 cm$^{-1}$ (a,b) and 2700-3100 cm$^{-1}$ (c,d) before (blue curves) and after (red curves) 400°C annealing. As-measured spectra are shown (i.e. without subtraction of a luminescent background and normalization to the diamond line intensity).

## 3. Conclusions

Ultrasmall nanodiamonds were synthesized at a pressure of 12 GPa and a temperature of about 1300$^0$C. A structural analysis of the nanomaterial was carried out using high-resolution transmission electron microscopy and Raman spectroscopy. The typical sizes of synthesized diamond crystallites were found to be 2-5 nm. The unusual vibration band containing a number of pronounced maxima at about 1147, 1245, 1344, and 1456 cm$^{-1}$ was detected in Raman spectra. The band is confidently identified with the bending vibrational modes of CHx groups terminating the nanodiamonds surface.

The 1000-1500 cm$^{-1}$ band is not detected either in 20-nm diamonds produced from adamantane [34] or in nanodiamonds >10 nm produced from halogenated adamantane [17]. Thus, the band is a distinctive spectral feature of ultrasmall nanodiamonds synthesized from adamantane and its derivatives. We suppose that the Raman band in the range of 1000-1500 cm$^{-1}$ can be used for express identification of UNDs synthesized not only from adamantane, but also from other hydrocarbons with a high hydrogen content. In turn, the identification of the spectral features of the ultra-small HPHT nanodiamonds with CHx bending modes confirms an extremely high hydrogen concentration on the surface of those diamond particles. This structural feature gives the green light to the use of the new generation of ultra-small nanodiamonds as a solid-state source of solvated electrons for heterogeneous catalysis [35] and radiotherapy [36,37]. Moreover, polarized CH bonds on a diamond surface are sensitive to environmental conditions. This opens up opportunities for using the diamond produced from adamantane as ultrasmall nanosensors in biology, chemistry, and medicine.

## 4. Methods

*Sample synthesis and preparation for analysis*

To create the 12 GPa pressure, a toroid type installation [38] was used. The profile of one of the two carbide anvils was a spherical recess surrounded by two toroidal grooves. The container, made of lithographic stone, had a profile on both sides, repeating the profile of the anvil. The graphite heater was placed in a cylindrical hole in its middle. Inside was a titanium container with adamantane. The experimental setup was pressure calibrated according to a well-known method [39]. Anomalies of conductivity of reference substances were recorded, which accompany phase transitions in reference substances with an increase in pressure at room temperature. For heating, an electric current was passed through the graphite heater. The temperature was recorded with thermocouples. In the diamond synthesis experiments the operating temperature was about $1300^0$C, the synthesis time - 1 min.

The diamond powder produced was dispersed in isopropanol, then a drop of the resulting suspension was deposited on a silicon substrate and dried. As a result, a thin island film consisting of diamond nanoparticles was formed on the substrate surface.

*Raman spectroscopy*

Raman spectra of the UND were recorded at room temperature under visible and UV laser excitations. For visible Raman records a LABRAM HR800 spectrometer and a 473-nm diode laser were used. UV Raman was recorded using a Triax 552 spectrograph, argon laser (514 nm) and WaveTrain CW frequency doubler (257 nm). The exciting laser radiation was focused into a spot of about 1 μm in diameter on the sample surface.

*Transmission electron Microscopy*

High resolution transmission electron microscopy (HRTEM) was performed using JEM ARM200F cold FEG double aberration corrected microscope, operated at 200 kV and equipped with a large angle CENTURIO EDX detector, OriusGatan CCD camera and Quantum GIF. The TEM samples were prepared in a conventional way – depositing of the sample from the suspension on a holey carbon supported copper grids.

*Sample annealing*

The substrate with the deposited sample was placed in the LinKam TS1500 chamber and heated in the air atmosphere three times, to temperatures of 200, 300, and 400°C. The heating rate was 120°C/minute. The sample was held at each of the three temperatures for 30 min, then the heating was turned off and the sample was cooled to room temperature. Raman spectra were recorded from the same area of the sample after each heating-cooling cycle.

**Funding:** This research was supported in part by Russian Science Foundation (grant No 22-19-00324). A.M.S acknowledges support from the RSF grant No. 22-21-00586.

**Acknowledgments:** We would like to thank Dr. A. Kirichenko for UV Raman analysis of our sample.

**Conflicts of Interest:** The authors declare no conflict of interest.